\documentclass[twocolumn,twocolappendix]{aastex7}

\usepackage{amsmath}

\usepackage{booktabs}   
\usepackage{multirow} 
\shorttitle{HST and GTC observations of AT2025ulz}
\shortauthors{Yang et al.}
\submitjournal{ApJL}

\usepackage{siunitx}
\usepackage{booktabs}
\usepackage{longtable}
\usepackage{caption}

\begin{document}

\title{
AT2025ulz and S250818k: zooming in with the Hubble Space Telescope
}

\correspondingauthor{Yu-Han Yang\\\href{mailto:yuhan.yang@roma2.infn.it}{yuhan.yang@roma2.infn.it}}
\author[0000-0003-0691-6688]{Yu-Han Yang}
\affiliation{Department of Physics, University of Rome ``Tor Vergata'', via della Ricerca Scientifica 1, I-00133 Rome, Italy}
\email{yuhan.yang@roma2.infn.it}

\author[0000-0002-1869-7817]{Eleonora Troja}
\affiliation{Department of Physics, University of Rome ``Tor Vergata'', via della Ricerca Scientifica 1, I-00133 Rome, Italy}
\affiliation{INAF - Istituto Nazionale di Astrofisica, 00133 Rome, Italy}
\email{eleonora.troja@uniroma2.it}

\author[0000-0001-7042-4472]{Marko Ristić}
\affiliation{Theoretical Division, Los Alamos National Laboratory, Los Alamos, NM 87545, USA}
\email{mristic@lanl.gov}

\author[0009-0004-9520-5822]{Muskan Yadav}
\affiliation{Department of Physics, University of Rome ``Tor Vergata'', via della Ricerca Scientifica 1, I-00133 Rome, Italy}
\email{muskan.yadav@students.uniroma2.eu}

\author[0009-0008-9010-2890]{Massine El Kabir}
\affiliation{Department of Physics, University of Rome La Sapienza, P.le Aldo Moro 2, 00185 Rome, Italy}
\affiliation{Department of Physics, University of Rome ``Tor Vergata'', via della Ricerca Scientifica 1, I-00133 Rome, Italy}
\email{massine.elkabir@uniroma1.it}

\author[0000-0002-7158-5099]{Rubén Sánchez-Ramírez}
\affiliation{Instituto de Astrof\'isica de Andaluc\'ia (IAA-CSIC), Glorieta de la Astronom\'ia s/n, 18008 Granada, Spain}
\affiliation{Unidad Asociada al CSIC Departamento de Ingeniería de Sistemas y Autom\' atica, Escuela de Ingenier\' ias Industriales, Universidad de M\'alaga, Arquitecto Francisco Pe\~nalosa, 6, 29071 M\'alaga, Spain}
\email{ruben@iaa.es}

\author[0000-0002-0216-3415]{Rosa L. Becerra}
\affiliation{Department of Physics, University of Rome ``Tor Vergata'', via della Ricerca Scientifica 1, I-00133 Rome, Italy}
\affiliation{Instituto de Astronom{\'\i}a, Universidad Nacional Aut\'onoma de M\'exico, Apartado Postal 70-264, 04510 M\'exico, CDMX, Mexico}
\email{rosa.becerra@roma2.infn.it}

\author[0000-0003-2624-0056]{Chris L. Fryer}
\affiliation{Theoretical Division, Los Alamos National Laboratory, Los Alamos, NM 87545, USA}
\affiliation{Center for Nonlinear Studies, Los Alamos National Laboratory, Los Alamos, NM 87545, USA}
\email{fryer@lanl.gov}

\author[0000-0002-9700-0036]{Brendan O'Connor}
\affiliation{McWilliams Center for Cosmology and Astrophysics, Department of Physics, Carnegie Mellon University, Pittsburgh, PA 15213, USA}
\email{boconno2@andrew.cmu.edu}

\author[0000-0001-6849-1270]{Simone Dichiara}
\affiliation{Department of Astronomy and Astrophysics, The Pennsylvania State University, 525 Davey Lab, University Park, PA 16802, USA}
\email{sbd5667@psu.edu}

\author[0000-0003-2999-3563]{Alberto J. Castro-Tirado}
\affiliation{Instituto de Astrof\'isica de Andaluc\'ia (IAA-CSIC), Glorieta de la Astronom\'ia s/n, 18008 Granada, Spain}
\affiliation{Ingeniería de Sistemas y Autom\'atica, Universidad de M\'alaga, Unidad Asociada al CSIC por el IAA, Escuela de Ingenier\'ias Industriales, Arquitecto Francisco Pe\~nalosa, 6, Campanillas, 29071 M\'alaga, Spain}
\email{ajct@iaa.es}

\author[0009-0002-6667-3294]{Camila Angulo-Valdez}
\affiliation{Instituto de Astronom{\'\i}a, Universidad Nacional Aut\'onoma de M\'exico, Apartado Postal 70-264, 04510 M\'exico, CDMX, Mexico}
\email{camiangulo@astro.unam.mx}

\author[0000-0002-6729-9022]{Josefa Becerra Gonz\'alez}
\affiliation{Instituto de Astrof\' isica de Andaluc\' ia (IAA-CSIC), Glorieta de la Astronom\' ia s/n, 18080 Granada, Spain}
\affiliation{Instituto de Astrofísica de Canarias and Universidad de La Laguna, Dpto. Astrofísica, 38200 La Laguna, Tenerife, Spain}
\email{jbecerra@iac.es}

\author[0000-0001-6650-2634]{Jos\'e A. Font}
\affiliation{Departamento de Astronom\'{\i}a y Astrof\'{\i}sica, Universitat de Val\`encia, Dr.~Moliner 50,  46100 Burjassot (Val\`encia), Spain}
\affiliation{Observatori Astron\`omic, Universitat de Val\`encia, Catedr\'atico Jos\'e Beltr\'an 2, 46980 Paterna (Val\`encia), Spain}
\email{j.antonio.font@uv.es}

\author[0000-0003-2238-1572]{Ori Fox}
\affiliation{Space Telescope Science Institute, 3700 San Martin Drive, Baltimore, MD 21218, USA}
\email{ofox@stsci.edu}

\author[0000-0001-7201-1938]{Lei Hu}
\affiliation{Department of Physics and Astronomy, University of Pennsylvania, 209 South 33rd Street, Philadelphia, PA, 19104, USA}
\affiliation{McWilliams Center for Cosmology and Astrophysics, Department of Physics, Carnegie Mellon University, Pittsburgh, PA 15213, USA}
\email{leihu@sas.upenn.edu}

\author[0000-0002-7400-4608]{Youdong Hu}
\affiliation{Guangxi Key Laboratory for Relativistic Astrophysics, School of Physical Science and Technology, Guangxi University, Nanning 530004, China}
\email{hyd@gxu.edu.cn}

\author[0000-0002-2467-5673]{William H. Lee}
\affiliation{Instituto de Astronom{\'\i}a, Universidad Nacional Aut\'onoma de M\'exico, Apartado Postal 70-264, 04510 M\'exico, CDMX, Mexico}
\email{wlee@astro.unam.mx}

\author[0000-0001-6148-6532]{Margarita Pereyra}
\affiliation{Secretar\'ia de Ciencia, Humanidades, Tecnolog\'ia e Innovaci\'on}
\affiliation{Instituto de Astronom\'ia, Universidad Nacional Aut\'onoma de M\'exico, A.P. 106, 22800 Ensenada, Baja California, M\'exico}
\email{mpereyra@astro.unam.mx}

\author[0000-0001-9050-7515]{Alicia M. Sintes}
\affiliation{Departament de Física, Universitat de les Illes Balears, IAC3 – IEEC, Carretera Valldemossa km 7.5, E-07122 Palma, Spain}
\email{alicia.sintes@uib.es}

\author[0000-0002-2008-6927]{Alan M. Watson}
\affiliation{Instituto de Astronom{\'\i}a, Universidad Nacional Aut\'onoma de M\'exico, Apartado Postal 70-264, 04510 M\'exico, CDMX, Mexico}
\email{alan@astro.unam.mx}

\author[0000-0002-9322-6900]{K. Océlotl C. López Mendoza}
\affiliation{Instituto de Astronom{\'\i}a, Universidad Nacional Aut\'onoma de M\'exico, Apartado Postal 70-264, 04510 M\'exico, CDMX, Mexico}
\email{koclopez@astro.unam.mx}

\begin{abstract}
AT2025ulz is an optical/near-infrared transient discovered during follow-up of the candidate gravitational wave (GW) event S250818k. Its young age ($\lesssim$1 d), rapid decline and strong color evolution over the first 48 hr classify it as a potential kilonova candidate.
In this work, we present the results of our observing campaign, carried out with the Gran Telescopio Canarias (GTC) and the \textit{Hubble Space Telescope} (\textit{HST}). 
Although the early time evolution of AT2025ulz resembles some aspects of a kilonova, its rapid onset ($\sim$3 hr after the GW trigger) and luminosity 
(a factor of $\sim5$ brighter than AT2017gfo in $g$-band) are difficult to reproduce. Only a small subset of our kilonova models matches its multi-color light curve, and the inferred ejecta mass is uncomfortably large given the low chirp mass ($\lesssim\!0.87\!$~M$_{\odot}$) of the GW candidate. 
\textit{HST} observations place the transient within a nearby ($z\!=\!0.08489$) spiral galaxy with on-going star-formation and measure a color ($F336W-F160W\!\approx\!1.4$ mag) that is too blue to match with a kilonova. 
Our data support the classification of AT2025ulz as a supernova, initially undergoing a shock-cooling phase and later entering its photospheric phase, and spectroscopically identified via its broad absorption features.
\end{abstract}


\section{Introduction} 
\label{sec:intro}
The first direct detection of gravitational waves (GWs) during the first LIGO observing run (O1) was a watershed moment in modern astrophysics \citep{GW150914}. 
Shortly thereafter, the second observing run (O2) led to the ground-breaking discovery of GW170817 \citep{GW170817, MMA}, a binary neutron-star (NS) merger detected in GWs and accompanied by a weak short gamma-ray burst (GRB 170817A; \citealt{Goldstein2017,Savchenko2017}), an optical/near-infrared (nIR) kilonova (AT2017gfo; \citealt{Coulter2017}) and a non-thermal afterglow radiation \citep{Troja2017,Hallinan2017}. 

Subsequent LIGO-Virgo-Kagra (LVK) observing runs \citep{Abbott2023GWTC3, LVK2025GWTC4Update} underscored the challenges of GW follow-up. GW localizations often extend over hundreds of square degrees with broad distance posteriors, requiring efficient search strategies that balance wide-field tiling and galaxy-targeted pointings informed by three-dimensional probability maps and galaxy catalogs \citep{Gehrels2016,Singer2016Distance,Evans2016,Arcavi2017,Coughlin2018,Ducoin2020,Salmon2020,Watson2020,Becerra21,GLADEplus2022,Cook2023,Coulter25}. 
Even as network sensitivities improve, a significant fraction of events retain large sky areas and poorly constrained distances \citep{Petrov22,Abbott20LRR}, complicating follow-up triage. Moreover, the local Universe is rich in unrelated transients: core-collapse supernovae (ccSNe) and Type Ia supernovae (SNe) dominate the optical transient rate and can masquerade as young, blue kilonovae if sampling is sparse \citep{Doctor17, Cowperthwaite2015}. 
Heterogeneous data streams (prediscovery limits, early colors, low-resolution spectra, radio/X-ray constraints) are rapidly combined to identify high-probability kilonova candidates and down-select interlopers \citep{Thakur2020,Vieira20,ENGRAVE20,GRANDMA20,Page2020,Paterson2021,ZTF24,Coulter25}.
Following this triage workflow, no event after AT2017gfo was elevated to the status of likely kilonova until AT2025ulz \citep{GCN41414, 2025GCN.41433....1H, 2025GCN.41452....1O} was discovered within the localization of the low-significance GW candidate S250818k \citep{GCN41437}.

S250818k\footnote{\url{https://gracedb.ligo.org/superevents/S250818k/view/}} is a sub-threshold GW candidate with false alarm rate (FAR) of $\approx$1 per five months. It was localized within an area of 786 deg$^2$ (90\% credible region) at a distance of 259$\pm$74~Mpc \citep{GCN41437}.
Despite its large localization area and high probability of being terrestrial noise (71\%), several follow-up campaigns were initiated.

Observations were motivated by the fact that since the start of O4, this event has the highest probability ($\approx$29\%) of being a binary neutron star (BNS) merger and thus to produce an electromagnetic signal.  

AT2025ulz was identified by \citet{GCN41414} at $r\!\approx\!21.3$ AB mag within a galaxy, classified as elliptical, at $z\!\approx\!0.09$, within the 3\,$\sigma$ distance interval set by the LVK collaboration. 
Attention to the source intensified when non-detections in ATLAS imaging constrained the rise to $\lesssim$1.7~days prior to discovery \citep{GCN41451}, strengthening the case for a very young transient. Within the next day, independent follow-up observations consistently measured a rapid decline in flux accompanied by pronounced reddening \citep{GCN41421,GCN41433,GCN41452,GCN41502}.
Spectroscopic observations determined the redshift more precisely to be $z=0.0848$ and revealed a smooth featureless spectrum consistent with a kilonova origin \citep{GCN41436}. 
Unfortunately, over the first four days, no near-infrared constraint was sufficiently sensitive to test this hypothesis \citep{GCN41456, GCN41476, GCN41504}. 

In this manuscript we report our observations of AT2025ulz with the \textit{Hubble Space Telescope} (\textit{HST}) which detected the source in the near-infrared and provided the first evidence of a color inconsistent with a kilonova interpretation. Moreover, our imaging provides a crisp view of its host galaxy, revealing a spiral morphology with on-going star-formation. 
We leverage our dataset with optical imaging and spectroscopy acquired with the Gran Telescopio Canarias (GTC). 
Our dataset supports the classification of this event as a Type II SN, unrelated to the GW candidate trigger.

Throughout the manuscript we adopt a standard $\Lambda$CDM cosmology \citep{Planck2020} with $H_0$\,$=$\,$67.4$ km s$^{-1}$ Mpc$^{-1}$, $\Omega_\textrm{m}$\,$=$\,$0.315$, and $\Omega_\Lambda$\,$=$\,$0.685$. 
All the errors and upper limits are reported at the 1$\sigma$ and 3$\sigma$ confidence level, respectively.

\begin{figure}
    \centering
    \includegraphics[width=0.45\textwidth]{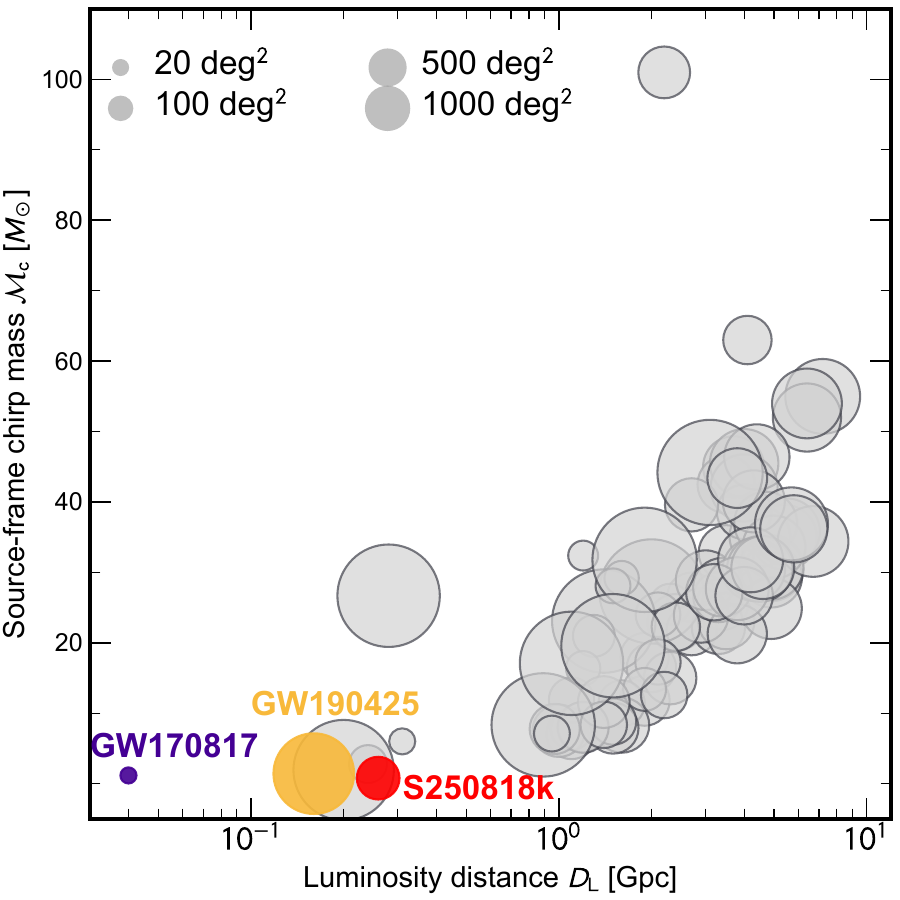} 
    \caption{
    Source-frame chirp mass ($\mathcal{M}_c$) v.s. luminosity distance ($D_{\rm L}$) for events detected during the O4a observing run. 
    Gray circles represent candidates of the \textit{GWTC-4.0} catalogue \citep{LVK2025GWTC4Update} with $\mathrm{FAR} \leq 1~\mathrm{yr^{-1}}$ and $P_{\mathrm{astro}} > 0.5$, where the circle size scales with the 90\,\% credible sky-localization area ($\Delta\Omega$). 
    } 
    \label{fig:O4a}
\end{figure}

\section{Observations}
\label{sec:obs}
Figure~\ref{fig:O4a} presents a summary of the GW events detected during the O4a observing run, together with the BNS mergers GW170817 \citep{GW170817} and GW190425 \citep{GW190425}. 
We use the \textit{GWTC--4.0} catalog \citep{LVK2025GWTC4Update} to sort the sources according to their luminosity distance ($D_{\rm L}$) and source-frame chirp mass ($\mathcal{M}_c$).
The circle size is proportional to the 90\,\% credible sky--localization area ($\Delta\Omega$). 

\begin{figure*}
	\includegraphics[clip, width=01\textwidth]{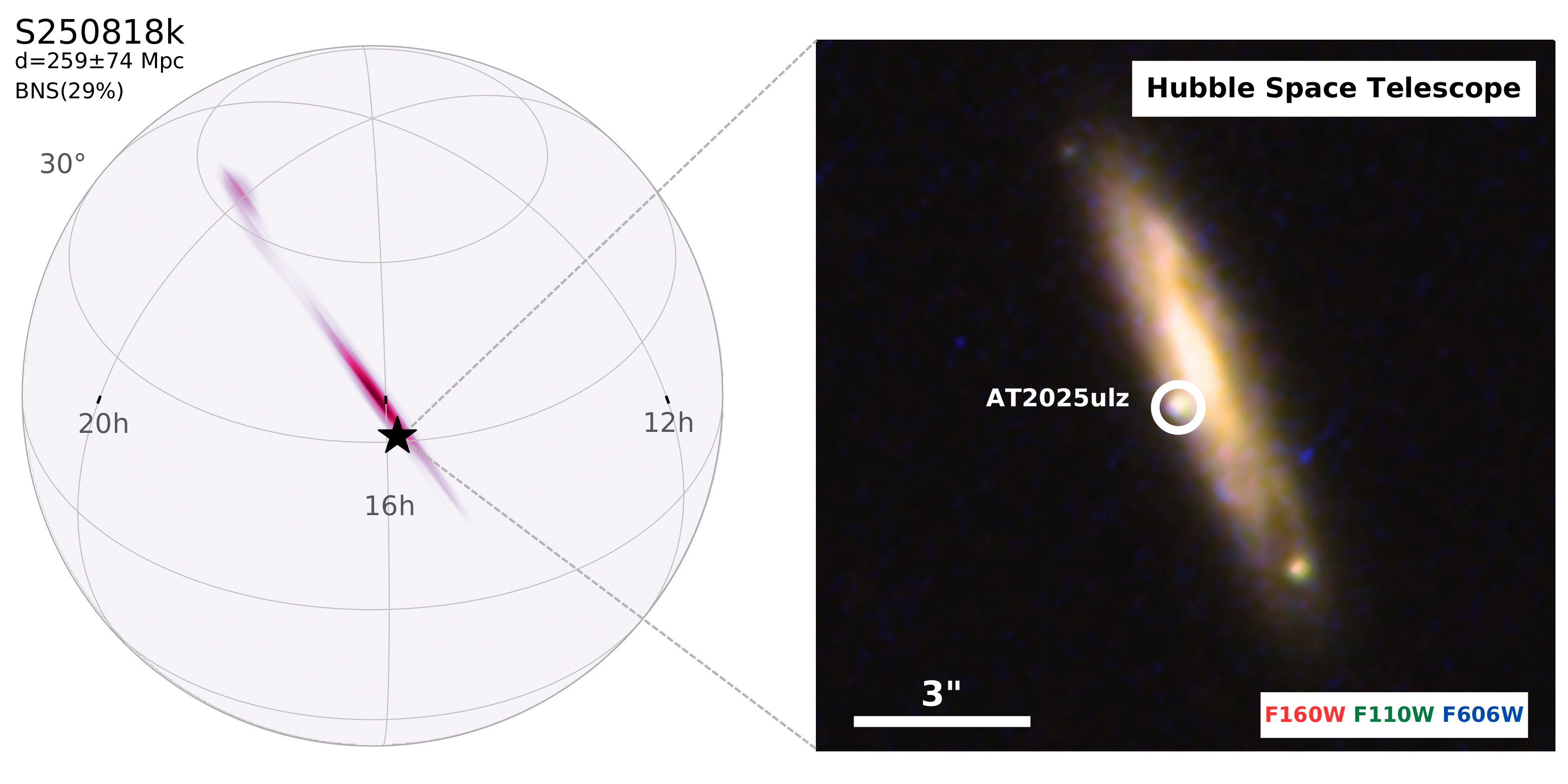}
    \caption{Probability map of the candidate GW event S250818k.
    A second lobe is located in the southern hemisphere and is not shown. 
    The position of AT2025ulz is marked by the star.
    A zoom-in view of the field from the \emph{Hubble Space Telescope} observations is displayed in the right panel, oriented with north up and east to the left.}
    \label{fig:field}
\end{figure*}

For S250818k, we used preliminary results from the Bayestar analysis reported by \citet{GCN41437}. The corresponding 90\,\% credible region spans $\Delta\Omega \sim 786~\mathrm{deg}^2$ and $D_{\rm L} = 259 \pm 74~\mathrm{Mpc}$ \citep{GCN41437}.
The plot shows that, despite the high probability of a noise event, in terms of mass, distance, and localization, this GW candidate represented a high-priority target for electromagnetic follow-up. 
Within the GW localization (Figure~\ref{fig:field}), \citet{GCN41414} reported the identification of over 50 transient sources and highlighted the discovery of AT2025ulz.

\subsection{Gran Telescopio Canarias}

We carried out the observations of AT2025ulz using the GTC with the upgraded Optical System for Imaging and low-Intermediate-Resolution Integrated Spectroscopy \citep[OSIRIS;][]{Cepa2000} under programs GTCMULTIPLE2A-25AMEX (PI: A. M. Watson), GTCMULTIPLE3A-25AMEX (PI: A. M. Watson) and GTCMULTIPLE3B-25A (PI: A. J. Castro Tirado).

\subsubsection{Imaging Observations}

Imaging observations were conducted on August 21 and 24, 2025 using the full set of $griz$ filters. The 1st epoch was observed with an average airmass of 1.2 and a seeing of 0.8\arcsec. The 2nd epoch was observed with an average airmass of 1.4 and a seeing of 0.8\arcsec. We also took the r-band acquisition image on August 30 with an airmass of 1.3 and seeing of 1\arcsec.

The photometry of the source is complicated by the bright host galaxy. 
We employed the saccadic Fast Fourier Transform (SFFT) algorithm \citep{Hu2022} to disentangle the transient signal from the host galaxy, derived from archival $grz$-band templates of the DESI Legacy Imaging Surveys \citep{Dey2019}\footnote{\url{https://www.legacysurvey.org/viewer}} for and $i$-band template of the Sloan Digital Sky Survey \citep[SDSS;][]{2022ApJS..259...35A}\footnote{\url{https://astroquery.readthedocs.io/en/latest/sdss/sdss.html}}.
Aperture photometry was performed using \texttt {Source Extractor} \citep{Bertin1996}, with results shown in Table~\ref{tab:gtc}. Magnitudes are calibrated against the Pan-STARRS DR2 catalog \citep{Flewelling2018} and in AB magnitude system \citep{Oke1983}.

\subsubsection{Spectroscopy}

Spectroscopic data were taken with a 0.8\arcsec~long slit with an average airmass of 1.4
and exposures times of $6\times300$~s and $5\times600$~s on August 21 and 30, 2025, respectively.
The blue region of the spectrum was observed with the R1000B grism, which provides a wavelength coverage of 3630--7500 \AA. The red part was covered with the R1000R grism with a range of 5100--10000 \AA.
In the second epoch, the slit was rotated by $\approx$90 degrees to minimize contamination by the galaxy's light. 
As shown in Figure \ref{fig:lcspec}a, several emission lines (H$\alpha$, H$\beta$, [OII], [OIII], [NII], [SII]) and absorption features (CaH\&K, NaID) are identified at a common redshift of $0.08489\pm0.00003$.

A comparison of the two spectra highlights the presence of a broad absorption feature at $\approx$6750 \AA\ in the latest spectrum, also identified in the VLT/MUSE  \citep{GCN41532} and Keck spectra \citep{GCN41538}
and interpreted as a P-Cygni H$\alpha$ profile, corresponding to an ejecta expansion velocity of $\approx$15,400 km s$^{-1}$. 
Broad undulations are visible at shorter wavelengths, although with lower significance. A dip at $\approx$5090 \AA\ could be consistent with a P-Cygni absorption profile from H$\beta$ for an expansion velocity of $\approx$10,300 km s$^{-1}$.

\subsection{Hubble Space Telescope}
\label{sec:HST}

We performed a Target of Opportunity (ToO) observation of AT2025ulz (GO17805; PI: E. Troja) starting on 2025-08-22 20:04:51 UT ($\approx$4.8 d since S250818k). 
Due to the reduced visibility of the target, \textit{HST} could only observe the field over a narrow window of less than 13~minutes per orbit. 
Therefore, we acquired short snapshots ($\approx$60--120 s) using the WFC3 camera with the UVIS channel and $F336W$ filter and the IR channel with the $F110W$ and $F160W$ filters. 

A second visit was performed under program GO17450 (PI: E. Troja) starting on 2025-08-26 21:32:20 UT acquiring short snapshots ($\approx$120--140~s per orbit) with the $F606W$ (UVIS), $F110W$, and $F160W$ (IR) filters. 

Calibrated images were downloaded from the Barbara A. Mikulski Archive for Space Telescopes (MAST), aligned using \texttt{Tweakreg} and combined  with \texttt{Astrodrizzle} \citep{Hoffmann2021} 
with pixfrac=0.7 to a final plate scale of 0.067\arcsec\ pixel$^{-1}$ for IR images, and 0.033\arcsec\ pixel$^{-1}$ for UVIS images. 

The galaxy contribution is significant in all our images, except that for the exposures in the $F336W$ filter. We used GALFIT \citep{Peng2002,Peng2010} to model the galaxy surface brightness (Sect.~\ref{sec:host}) and subtracted the best-fit model from each drizzled image to isolate point-like residuals.
Aperture photometry was then performed on the residual images using circular apertures of radius  0.15\arcsec and 0.16\arcsec\ for the IR and UVIS exposures, respectively. The derived fluxes were corrected for PSF losses using the tabulated encircled-energy values\footnote{\href{https://www.stsci.edu/hst/instrumentation/wfc3/data-analysis/photometric-calibration/ir-encircled-energy}{https://www.stsci.edu/hst/instrumentation/wfc3}} and converted to AB magnitudes using the appropriate zeropoint recorded in the image headers.

\section{Results}
\label{sec:results}

\subsection{Temporal Analysis}

\begin{figure*}
   \centering
   \includegraphics[width=1\linewidth]{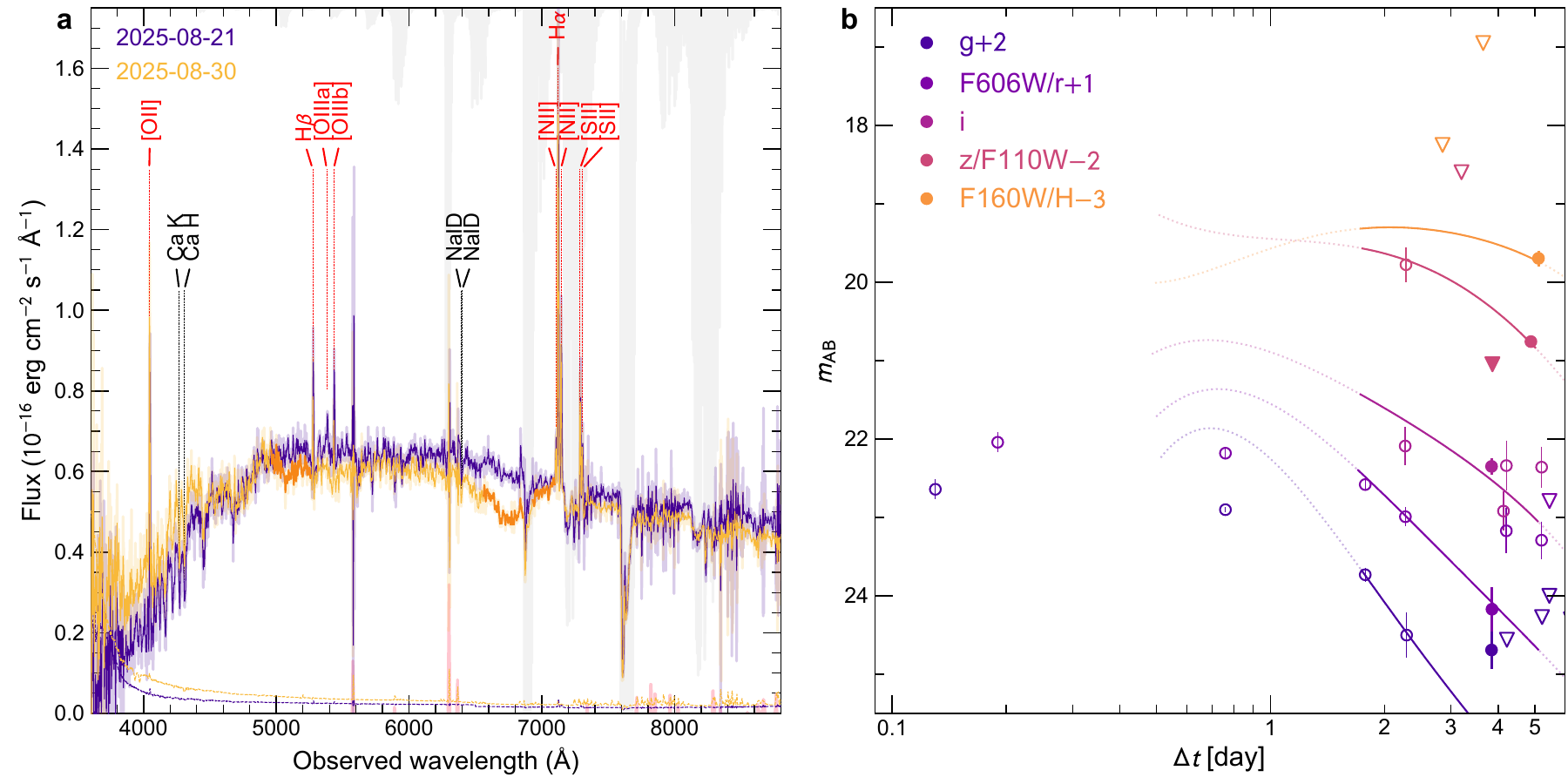}
   \caption{ GTC spectra (\textbf{a}) and multi-color light curve (\textbf{b}) of AT2025ulz, corrected for Galactic extinction.
   \textbf{a}. The observed spectra are plotted as semi-transparent lines, with smoothed solid lines for plotting purposes and dashed lines for uncertainties. 
   Both spectra contain significant, though different, levels of host-galaxy contribution. The second-epoch spectrum, which is less affected by the host, has been rescaled by a factor of two for comparison.
   Emission and absorption lines are marked by the red and black dashed lines, respectively. 
   The P-Cygni profiles of H$\alpha$ and H$\beta$ are highlighted in a deeper color shade.
   Wavelengths affected by telluric absorption and sky emission are marked by gray and pink areas, respectively \citep{Lord1992}.
   \textbf{b}. Circles and downward triangles denote detections and upper limits, respectively. Filled symbols indicate our observations (Table \ref{tab:gtc}), while open symbols correspond to supplementary data (Table \ref{tab:gcn} and \citealt{Gillanders2025}). 
   The data are scaled for plotting purposes. 
   The smoothed light curves of AT2017gfo, redshifted to $z = 0.08489$ and further shifted by $\Delta$m $= -1.8$ ($g$), $-1.5$ ($r$), $-1.3$ ($i$),  $-1$ ($z$), $-0.2$ ($H$), are shown as dotted lines for comparison with AT2025ulz. The similar trend of evolution is highlighted by the solid lines.
   }
   \label{fig:lcspec}
\end{figure*}

We built a multi-color light curve, shown in Figure \ref{fig:lcspec}b, using the data in Table \ref{tab:gtc} and supplementary data from Table \ref{tab:gcn} and \citet{Gillanders2025}. 
The observed magnitudes were corrected for Galactic extinction by adopting the extinction law from \cite{Gordon2023}\footnote{Implemented as the G23 model in the Python package \texttt{dust\_extinction}, which builds upon results from \citet{Gordon2009}, \citet{Fitzpatrick2019}, \citet{Gordon2021}, \citet{Decleir2022}.} with $R_V=3.1$ and $A_V=0.29$, derived from the \textit{Gaia} Total Galactic Extinction (TGE) map\footnote{We queried the highest pixel resolution available for the coordinates in the TGE map, i.e., healpix\_level $= 7$.} \citep{Delchambre2023}. 

The early ($\lesssim$4 d) light curve of AT2025ulz displays a rapid decay and hints at a chromatic evolution, with the bluer filters decaying faster than the redder ones. 
During the same time interval, the kilonova AT2017gfo displays a remarkably similar behavior. This is shown in Figure \ref{fig:lcspec}b, where we collected data from the literature \citep{2017ApJ...848L..17C, 2017ApJ...848L..24V, 2017ApJ...848L..27T, 2017ApJ...848L..29D, 2017Natur.551...64A, 2017Natur.551...67P, Troja2017, 2017Natur.551...75S} to build a smoothed light curve of AT2017gfo and rescaled it to match the observed photometry of AT2025ulz. 
Therefore, based on the early-time temporal and color evolution, AT2025ulz resembled the behavior of a kilonova, at least in the optical range.  However, its luminosity exceeded that of AT2017gfo at a similar epoch ($\Delta m_r\!\approx\!-1.5$ mag).

 A kilonova classification is not confirmed by
 observations at later time ($\gtrsim$4 d). 
 Our \textit{HST} imaging detects the source in the nIR filters and in the blue $F336W$ filter at 4.8 d. 
The observed color $F336W-F160W\!\approx\!1.4$ mag is substantially different from the color of AT2017gfo at a similar epoch \citep[$F336W-F160W\!\approx\!7$ mag;][]{Kasliwal2017,Troja2017}. 
Moreover, the observed flux is seen to rise out to 12.8~d in $r$-band when our monitoring ended. 
The color, rising light curve, and luminosity are instead naturally explained by a SN origin.

\subsection{Host Galaxy}
\label{sec:host}
We performed a two–dimensional morphological analysis of the host galaxy with GALFIT \citep{Peng2002,Peng2010}, using our $HST$ images in the $F606W$ and $F160W$ filters.  
We modeled the galaxy's surface brightness using a single S\'ersic profile,  parametrized with an effective radius $R_e$, index $n$, axis ratio $q\!=\!b/a$, and 
position angle PA. GALFIT convolves the model with the point spread function (PSF), selected from a grid of spatially-dependent models \citep{AndersonKing2006},
and compares it to the observed surface brightness. 
Model performance was evaluated using $\chi^2$ minimization and inspection of the residual images, obtained by subtracting the best fit model from the science images.

The single component model leaves prominent residuals, indicating a more complex morphology. 
The best fit is obtained by combining two S\'ersic profiles. 
The brighter component ($F606W\approx18.7$ mag) with $n\!\approx\!1.1$,
$R_e\!\approx\!2.8$\arcsec, $q\!\approx\!0.24$, and PA\,$\!\approx\!60^{\circ}$ is consistent with an exponential disk. 
The fainter component ($F606W\approx19.1$ mag) has similar best fit parameters but $n\!\approx\!0.17$ and PA\,$\!\approx\!65^{\circ}$,
approximating tilted spiral arm structures rather than a central bulge. 
Therefore, the initial classification of the host galaxy as elliptical is not consistent with our imaging analysis. Instead, the system looks as a spiral galaxy (Figure~\ref{fig:field}) with no obvious bulge structure and seen nearly edge-on. Because the inclination is high, individual arms are hard to trace and the exact sub-type remains uncertain. 

This classification is consistent with the spectral results, featuring bright nebular lines from on-going star-formation alongside absorption features from the stellar population and a dusty interstellar medium. From line fitting we derive a Balmer decrement of ${\rm H}\alpha/{\rm H}\beta\approx\!4.5$ and a luminosity $L_{{\rm H}\alpha}\!\approx\!1.3\times10^{40}$\,erg s$^{-1}$ from which we estimate a star formation rate SFR\,$\approx0.2$\,$M_{\odot}$\,yr$^{-1}$ using standard relations \citep{Kennicutt2012, Chabrier2003, Cardelli1989}. The UV emission  \citep[$M_{\rm FUV}\approx21.7$ mag;][]{Bianchi2017} combined with the radio continuum \citep[$F_{\rm 1.3GHz}\approx186\ \mu$Jy and spectral index $\alpha\sim-0.9$;][]{Franz2025} yields SFR$\approx\!1.3\,M_{\odot}$\,yr$^{-1}$ \citep{Kennicutt2012,Hao2011}. This difference may arise from potential slit–loss bias in the optical measurement, from dust–obscured star formation not captured by optical diagnostics, or from a recent decline in the SFR given the different tracer timescales \citep[see also][]{Hall2025a}.

From the $F160W$ image, we measure a galactocentric offset of 0.88\arcsec\,$\pm$\,0.04\arcsec, corresponding to a projected physical offset of $1.45\pm0.06$ kpc. 
The host-normalized offset is $\approx$0.35. 
These values are rather small compared to the population of high-energy transients \citep{OConnor2022, Anderson2015, Fong2013, Kelly2008, Fruchter2006, Bloom2002} and given the high inclination of the system and the compression of lengths along the line of sight, they should be considered as lower bounds to the true offset.

\begin{figure*}
    \centering
    \includegraphics[width=1\linewidth]{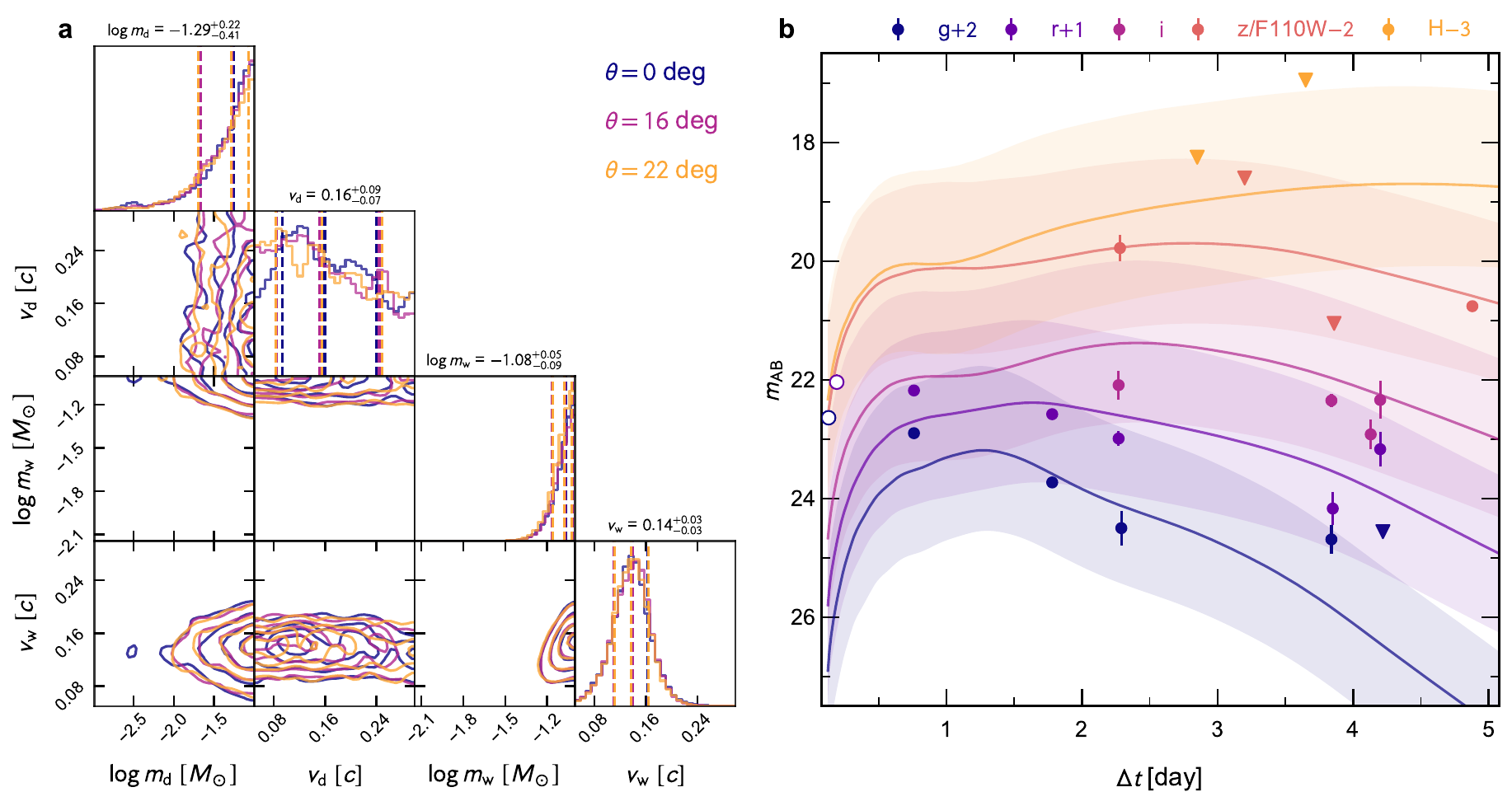}
    \caption{Posteriors of parameters (\textbf{a}) and related multi-wavelength light curve with viewing angle $\theta=0$~deg (\textbf{b}) of kilonova model. Data used for the kilonova model fit are plotted as filled circles. Early ($\approx$3-4 hr) detections of AT2025ulz (white circles) cannot be reproduced by our kilonova models.
    \label{fig:inference}}
\end{figure*}

\subsection{Constraints on Kilonova Model}

We compare the early-time ($<5$ day) detections (not upper limits) from Tables~\ref{tab:gtc} and \ref{tab:gcn} to the publicly available two-component kilonova models from \citet{wollaeger_2021_7335961}, specifically the subset presented in \citet{2022PhRvR...4a3046R}. 
Our model assumes an equatorially symmetric, lanthanide-rich, toroidal dynamical ejecta component and an axially symmetric, lanthanide-poor, double-lobed wind component.
These simulated kilonova spectra were generated using \texttt{SuperNu} \citep{SuperNu}, a Monte Carlo code for simulation of time-dependent radiation transport with matter in local thermodynamic equilibrium \citep[see, e.g.,][for further description of \texttt{SuperNu} physics assumptions]{2021ApJ...918...10W,2022PhRvR...4a3046R}.

We train a \texttt{scikit-learn} \citep{scikit-learn} Gaussian process (GP) on 412 two-component simulations with each simulation parameterized by inputs $\vec{x} = [m_{\rm{d}}, v_{\rm{d}}, m_{\rm{w}}, v_{\rm{w}}]$, the dynamical and wind ejecta masses and velocities, respectively, and outputs $y_{\rm{b}} = M_{\rm{b}}(\vec{x})$, where $M_b$ are the absolute magnitudes for a given broadband filter $b$.
Our GP smoothly interpolates over $\vec{x}$ within our grid's parameter limits $0.001 < m/M_\odot < 0.1$ and $0.05 < v/c < 0.3$ to provide a continuous representation $\hat{y}_{\rm{b}}(\vec{x})$ for arbitrary inputs $\vec{x}$. We use this GP for Bayesian inference with a purely Gaussian likelihood which takes the form
\begin{equation}
  \ln {\cal L} = -\frac{1}{2}  \sum_{t, B}\left[ \frac{(\hat{y}_{t, B} - d_{t, B})^2}{\sigma_t^2 + \sigma_{\rm sys}^2} + \ln [2\pi(\sigma_t^2+\sigma_{\rm sys}^2)] \right] \,,  
\end{equation}
where $\hat{y}_{t, B}$ is our GP prediction at time $t$ in band $B$ for input $\vec{x}$, $d_{t, B}$ are observations at the same time and band, and $\sigma_t$ and $\sigma_{\rm{sys}}$ are the observational and systematic uncertainty, respectively. In our analysis, we treat $\sigma_{\rm{sys}}$ as a free parameter during inference, recovering values $\sim 1$ mag.

The posterior distributions of the ejecta parameters and fitted light curves are shown in Figure \ref{fig:inference}.
We consider several viewing angle bins ($0^\circ \leq \theta < 16^\circ$, $16^\circ \leq \theta < 22^\circ$, $22^\circ \leq \theta < 27^\circ$) during inference, but are unable to recover viewing angle dependencies, likely owing to the large systematic uncertainties required to fit the data ($\sim 1$ mag).
We recover large ejecta component masses of $m_{\rm{d}} = 0.05 M_\odot$ and $m_{\rm{w}} = 0.08 M_\odot$, both of which approach the upper mass limit of $0.1 M_\odot$ in our baseline simulation grid.
The total ejecta mass of the system is $m_{\rm{ej}} \approx 0.13 M_\odot$.

This result is at odds with a BNS origin for S250818k, characterized by a relatively low
chirp mass $\mathcal{M}_{\rm{c}} \sim 0.1-0.87\ M_\odot$ \citep{GCN41440}. 
Chirp mass values below 0.96 $M_\odot$ are challenging to achieve even in the most extreme neutron star formation scenarios ~\citep{Fryer1999,Cheong2025}.  
Although an equal-mass BNS composed of two 1.0\,M$_\odot$ neutron stars could produce a chirp mass of 0.87 $M_\odot$, such systems are expected to eject very little dynamical or wind mass~\citep[$\lesssim 0.01\ M_\odot$; e.g.,][]{2021ApJ...906...98N, 2023PhRvD.107f3028H}, insufficient to account for the ejecta mass implied by AT2025ulz.

Moreover, radioactively powered kilonova models cannot naturally account for the bright optical emission detected a few hours after the merger \citep{2021ApJ...913..100K, 2022PhRvR...4a3046R}.
These models can display variability in the optical at early times depending upon the velocity distribution~\citep{2024ApJ...961....9F}, but this variability is limited to emission within 1\,d of the merger.
Energy injection from a central engine is unlikely to act on such short timescales as it must propagate outward \citep{2019ApJ...880...22W}.  Models incorporating ejecta cocoon emission could alleviate the mismatch with the early optical lightcurve \citep[e.g.][]{2018MNRAS.479..588G, 2018ApJ...855..103P}, though the timescale of sufficiently bright cocoon emission remains inconclusive \citep[see, e.g.,][]{2023MNRAS.524.4841H}. Jet-ejecta interaction can potentially increase the early UV and optical emission, although some models 
\citep[e.g.][]{2021MNRAS.500.1772N}
produce a peak only after 1\,d while others are dominated by $r$-process heating \citep{2021MNRAS.502..865K}. In either case, depending on the observer's angle, successful jets should lead to (possibly delayed) radio and X-ray emission which was not observed for this event \citep{OConnor2025ulz}.

\begin{figure}
    \centering
    \includegraphics[width=1\linewidth]{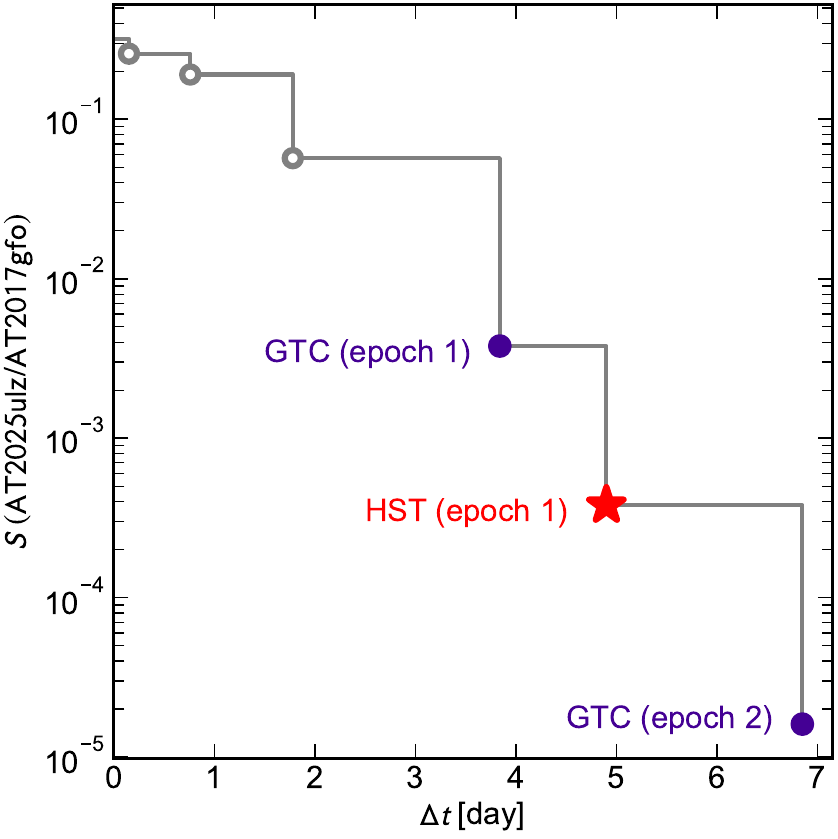}
    \caption{
    Temporal evolution of the score of AT2025ulz being a kilonova with respect to AT2017gfo at the same time.}
    \label{fig:pkn}
\end{figure}

\subsection{Classification}

The classification of AT2025ulz evolved rapidly as the dataset grew.
Several criteria must be considered when trying to identify a kilonova among a myriad of other transients: sky position consistent with the GW signal
(Figure~\ref{fig:field}), young age estimated from historical data, luminosity compatible with a kilonova
(Figure~\ref{fig:inference}) and a spectrum characterized by red color ($g-i$\,$>$0). Additionally, one has to factor in the probability (as released by the LVK collaboration) that the GW merger contains at least a NS.
The properties of the host galaxy as well as the location of the transient within the
host are also considered in the decision tree (Section~\ref{sec:host}).
Based on the above criteria (see Appendix Section~\ref{sec:rank}), we compare the ranking of AT2025ulz, derived from our dataset, with the ranking of the kilonova AT2017gfo at a similar epoch.
Figure~\ref{fig:pkn} shows that, prior to any optical rise or spectral absorption, 
the color and host environment inferred from \textit{HST} substantially reduced the likelihood of a kilonova.

\section{Conclusions}
\label{sec:conclusions}

The sub-threshold GW event S250818k, with its probability of being a binary NS merger, coupled with the optical transient AT2025ulz identified within a nearby and possibly elliptical galaxy \citep{GCN41414}, triggered a broad interest for further observations across multiple facilities.

\textit{HST} imaging reveals that the transient was located in a nearly edge-on spiral galaxy, where star-formation is still on-going. Such environment is consistent with a broad range of stellar progenitors, from NS mergers to massive stars.

Our analysis shows that, at early times, some of the properties of AT2025ulz are reminiscent of a kilonova, at least in the optical band. 
However, several other ones, such as its early peak ($\approx$~3 hr) and color ($F336W-F160W\!\approx\!$~1.4 mag at 4.8 d), are not consistent with this interpretation. 
Early peak emission is not reproduced by any of our kilonova models, and is more indicative of SN shock breakout and cooling.  
The strength of the supernova cooling emission depends both on internal shocks and shocks in the immediate circumstellar material, but typically is characterized by strong UV emission with cooling timescales of 1--10\,d~\citep{2017ApJ...850..133D,2022ApJ...931...15B,2025ApJ...990L..68S}.  Ultraviolet emission from kilonova \citep{Evans2017} 
tends to drop off more quickly ($<\!1$d) than supernova shock cooling~\citep{2024ApJ...961....9F}.  Strong UV after 1\,d, as observed with \textit{HST}, indicates a SN shock cooling event.  

After the cooling peak, SN models predict a rise as the standard SN light curve dominates the emission. Therefore, a SN naturally explains the late-time ($\gtrsim$5 d) evolution of AT2025ulz. Broad absorption features in our GTC optical spectrum confirm that the emission was powered by fast-moving ejecta. Their identification as P-Cygni profiles in the Balmer series indicates an ejecta composition (H-rich)  and velocity ($\approx$10,000--16,000 km~s$^{-1}$) characteristic of a Type II SN.

\begin{acknowledgments}
This work is supported by the European Research Council through the Consolidator grant BHianca (grant agreement ID~101002761).

The work by CLF and MR is supported at Los Alamos National Laboratory (LANL) which is operated by Triad National Security, LLC, for the National Nuclear Security Administration of U.S. Department of Energy (Contract No. 89233218CNA000001). This document has been assigned LA-UR-25-31126.
MR acknowledges support from the Information Science \& Technology Institute (ISTI) at LANL as an ISTI Postdoctoral Fellow and acknowledges PECASE award funds. AMW is grateful for support from UNAM/DGAPA PAPIIT project IN109224. CAV acknowledges support from a SECIHTI fellowship. BO is supported by the McWilliams Postdoctoral Fellowship in the McWilliams Center for Cosmology and Astrophysics at Carnegie Mellon University. 
JBG acknowledges support from the Agencia Estatal de Investigación del Ministerio de Ciencia, Innovación y Universidades (MCIU/AEI) under grant PARTICIPACIÓN DEL IAC EN EL EXPERIMENTO AMS and the European Regional Development Fund (ERDF) with reference PID2022-137810NB-C22.
AMS is supported by the Spanish Agencia Estatal de Investigación grants PID2022-138626NB-I00, RED2024-153978-E, RED2024-153735-E, funded by MICIU/AEI/10.13039/501100011033 and the ERDF/EU; and the Comunitat Autònoma de les Illes Balears through the Conselleria d'Educació i Universitats with funds from the European Union - NextGenerationEU/PRTR-C17.I1 (SINCO2022/6719) and from the European Union - European Regional Development Fund (ERDF) (SINCO2022/18146).
AJCT acknowledges support from the Spanish Ministry project PID2023-151905OB-I00 and Junta de Andaluc\'ia grant P20\_010168 and from the Severo Ochoa grant CEX2021-001131-S funded by MCIN/AEI/ 10.13039/501100011033.
JAF is supported by the Spanish Agencia Estatal de Investigaci\'on (grants PID2021-125485NB-C21 and PID2024-159689NB-C21) funded by MCIN/AEI/10.13039/501100011033 and ERDF A way of making Europe, by the Generalitat Valenciana (Prometeo grant CIPROM/2022/49), and by the  European Horizon Europe staff exchange (SE) programme HORIZON-MSCA-2021-SE-01 (NewFunFiCO-101086251).

This work is partly based on observations with the NASA/ESA \textit{Hubble Space Telescope} obtained from the Mikulski Archive for Space Telescopes (MAST) hosted at the Space Telescope Science Institute, which is operated by the Association of Universities for Research in Astronomy, Incorporated, under NASA contract NAS526555.
The material is based upon work supported by the Space Telescope Science Institute under award number HST-GO-17805.002-A.
The HST data presented in this article were obtained from the MAST at the Space Telescope Science Institute. The specific observations analyzed can be accessed via \dataset[doi: 10.17909/h3df-ay84]{https://doi.org/10.17909/h3df-ay84}.

This work is also partly based on data obtained with the Gran Telescopio Canarias (GTC), installed in the Spanish Observatorio del Roque de los Muchachos of the Instituto de Astrofísica de Canarias, in the island of La Palma, and with the instrument OSIRIS, built by a Consortium led by the Instituto de Astrofísica de Canarias in collaboration with the Instituto de Astronomía of the Universidad Autónoma de México. OSIRIS was funded by GRANTECAN and the National Plan of Astronomy and Astrophysics of the Spanish Government. 

We thank the staff of HST and GTC for scheduling and executing the observations included in this work.

Sky-background data is provided by \texttt{ATRAN} \citep{Lord1992} and Gemini Observatory.
\end{acknowledgments}

\facilities{
\textit{Hubble Space Telescope}, 
\textit{Gran Telescopio Canarias}
}

\software{\href{https://www.astropy.org}{Astropy} \citep{astropy:2013, astropy:2018, astropy:2022}, 
\href{https://matplotlib.org}{Matplotlib} \citep{matplotlib}, 
\href{https://scipy.org/}{SciPy} \citep{Virtanen2020}, 
\href{https://numpy.org/}{NumPy} \citep{numpy},
\href{https://pandas.pydata.org/}{pandas} \citep{pandas},
\href{https://scikit-learn.org/stable/}{scikit-learn} \citep{scikit-learn}, \href{https://corner.readthedocs.io/en/latest/}{corner} \citep{corner}, 
IRAF \citep{Tody1986,Tody1993}, \href{https://users.obs.carnegiescience.edu/peng/work/galfit/galfit.html}{GALFIT} \citep{Peng2002,Peng2010}, \href{https://drizzlepac.readthedocs.io/en/latest/}{DrizzlePac} \citep{Hoffmann2021}, \href{https://www.astromatic.net/software/sextractor/}{Source Extractor} \citep{Bertin1996}, \href{https://github.com/thomasvrussell/sfft}{SFFT} \citep{Hu2022}, \href{https://github.com/lanl/SuperNu}{SuperNu} \citep{SuperNu}, \href{https://sites.google.com/cfa.harvard.edu/saoimageds9}{SAOImageDS9} \citep{Joye2003}, \href{https://dustmaps.readthedocs.io/en/latest/}{dustmaps}
} \citep{dustmaps},
\href{https://dust-extinction.readthedocs.io/en/latest/}{dust\_extinction} \citep{dust_extinction}
\href{https://lscsoft.docs.ligo.org/ligo.skymap}{ligo.skymap}\citep{Singer2016a,Singer2016Distance,Singer2016c}

\appendix
\renewcommand{\thetable}{A\arabic{table}}
\renewcommand{\thefigure}{A\arabic{figure}}
\setcounter{table}{0}
\setcounter{figure}{0}

\section{Log of observations}
\begin{deluxetable*}{ccccccc}
\setlength{\tabcolsep}{12pt}
\tablecaption{GTC and HST observations log. $\Delta t$ is the difference of the mid-time of the observation and the GW trigger time 2025-08-18 01:20:06.030 UT. Magnitudes are not corrected for Galactic extinction.
\label{tab:gtc}}
\tablehead{
\colhead{Start Date}&
\colhead{$\Delta t$}&
\colhead{Telescope}& 
\colhead{Instrument}&
\colhead{Filter}& 
\colhead{Exposure}& 
\colhead{AB Mag}\\
(UTC)
&\colhead{(d)}&
&&&
\colhead{(s)}&
}
\startdata
2025-08-21 21:23:47&3.84&GTC&OSIRIS&$g$&120&$23.04 \pm 0.24$\\
2025-08-21 21:25:56&3.84&GTC&OSIRIS&$i$&120&$22.53 \pm 0.11$\\
2025-08-21 21:40:25&3.85&GTC&OSIRIS&$r$&480&$23.42 \pm 0.28$\\ 
2025-08-21 21:53:36&3.86&GTC&OSIRIS&$z$&480&$>23.15$\\
2025-08-22 20:04:51 &4.78 & HST & WFC3/UVIS & F336W &$120$&  $24.15\pm0.15$  \\
2025-08-22 21:39:37 &4.88 & HST & WFC3/IR & F110W &$127$& $22.85\pm0.05$  \\
2025-08-23 00:48:14 &5.11 & HST & WFC3/IR & F160W &$127$& $22.75\pm0.10$  \\
2025-08-24 21:18:21&6.84&GTC&OSIRIS&$g$&1210&$22.43\pm0.14$\\ 
2025-08-24 21:49:51&6.86&GTC&OSIRIS&$r$&960&$22.23 \pm 0.09$\\ 
2025-08-24 22:14:00&6.88&GTC&OSIRIS&$i$&540&$22.08 \pm 0.07$\\ 
2025-08-24 22:27:38&6.89&GTC&OSIRIS&$z$&600&$22.32\pm0.23$\\ 
2025-08-26 21:32:20 &8.84 & HST & WFC3/UVIS & F606W &$120$&  $21.78\pm0.02$  \\
2025-08-27 19:33:22 &9.76 & HST & WFC3/IR & F110W &$143.8$&   $21.97\pm0.03$\\
2025-08-28 11:17:11 &10.42 & HST & WFC3/IR & F160W &$143.8$& $22.26\pm0.04$  \\
2025-08-30 21:17:44 &12.83&GTC&OSIRIS&$r$&60&$21.20 \pm 0.04$\\
\enddata
\end{deluxetable*}
\begin{table}[h!]
\centering
\begin{tabular}{ccccc}
\hline
\hline
$\Delta t$ & Telescope & Filter & Magnitude & Reference \\
\hline
$-$0.84 & ZTF & $r$ & $> 21$ & (1) \\
0.13 & ZTF & $g$ & $20.99 \pm 0.13$ & (1) \\
0.19 & ZTF & $r$ & $21.29 \pm 0.13$ & (1) \\
0.76 & FTW & $g$ & $21.25 \pm 0.03$ & (2) \\
0.76 & FTW & $r$ & $21.43 \pm 0.06$ & (2) \\
1.08 & WINTER & $J$ & $> 19.30$ & (3) \\
1.78 & FTW & $g$ & $22.08 \pm 0.09$ & (4) \\
1.78 & FTW & $r$ & $21.83 \pm 0.06$ & (4) \\
2.8 & TNG & $J$ & $> 21.70$ & (5) \\
2.85 & TNG & $H$ & $> 21.30$ & (5) \\
2.9 & VLT & $K$ & $> 22.00$ & (6) \\
2.9 & TNG & $K$ & $> 20.50$ & (5) \\
3.65 & PRIME & $H$ & $> 20.00$ & (7) \\
4.13 & COLIBRÍ  & $i$ & $23.10 \pm 0.25$ & (8) \\
6.14 & COLIBRÍ & $i$ & $22.00 \pm 0.10$ & (8) \\
\hline
\end{tabular}
\caption{Photometric observations of AT2025ulz from public resources. Magnitudes are not corrected for Galactic extinction. References: 1. \citealt{2025TNSTR3264....1S}, 2. \citealt{GCN41421}, 3. \citealt{GCN41456}, 4. \citealt{GCN41433}, 5. \citealt{GCN41489}, 6. \citealt{GCN41476}, 7. \citealt{GCN41504}, 8. \citealt{GCN41518}.\label{tab:gcn}}
\end{table}

Table \ref{tab:gtc} summarizes the results of imaging observations obtained in this work, while Table \ref{tab:gcn} provides supplementary photometric data collected from public databases.

\section{Ranking of AT2025ulz}
\label{sec:rank}

To ensure efficient allocation of resources, our analysis seeks to establish whether a GW event and its candidate counterpart merit observational follow-up. 
Potential counterparts are assigned a score, defined as: 
\begin{equation}
    S = P_{\rm NS}\times S_{\rm 3D} \times S_{\rm age} \times S_{\rm KN} \times P_{\rm galaxy},
\end{equation}
where $P_{\rm NS}$ estimates the probability of an astrophysical event involving at least a NS, $S_{\rm 3D}$ the positional consistency,   $S_{\rm age}$ ranks the counterpart based on its age,  $S_{\rm KN}$  and $P_{\rm galaxy}$ assess whether the candidate counterpart is a kilonova based on the observed photometry and its galaxy association and morphology, respectively.  
We take GW170817 and its counterpart AT2017gfo as a benchmark against which to compare our candidates.

Our assessment begins with the probability that the GW signal is of astrophysical nature and involves at least one NS, a prerequisite for electromagnetic emission. 
The relative probability of S250818k is a genuine BNS merger is $P_{\rm NS}\left(\tfrac{\rm S250818k}{\rm GW170817 }\right)=0.29$.

The three-dimensional (3D) consistency of the galaxy–candidate association in sky position and distance is inferred from the 3D GW posterior probability
distribution \citep{Singer2016Distance,Singer2016ApJS}. We cross-match the GLADE+ catalog \citep{2022GLADE+} with the posterior sky map and rank all potential host galaxies according to the 3D probability density (Mpc$^{-3}$) within the 99\% credible volume\footnote{\url{https://lscsoft.docs.ligo.org/ligo.skymap/postprocess/crossmatch.html}}. 
Given the low rate of BNS mergers, the correspondingly high observational interest in retaining all plausible EM counterparts, and the sharp differences in the 3D posterior probability density across the localization volume, we adopt a conservative unweighted rank–percentile score, defined as
\begin{equation}
    S_{\rm 3D}(k) = 1-\frac{k-1}{N},
\end{equation}
where $N$ is the number of candidate hosts within the localization volume and $k$ is the rank of a given galaxy in descending posterior density.
The resulting $S_{\rm 3D}\left(\tfrac{\rm AT2025ulz}{\rm AT2017gfo }\right) = 0.69$ with $S_{\rm 3D}({\rm AT2025ulz}) = 0.66$ and $S_{\rm 3D}({\rm AT2017gfo}) = 0.96$.

\begin{figure*}
    \centering
    \includegraphics[width=1\linewidth]{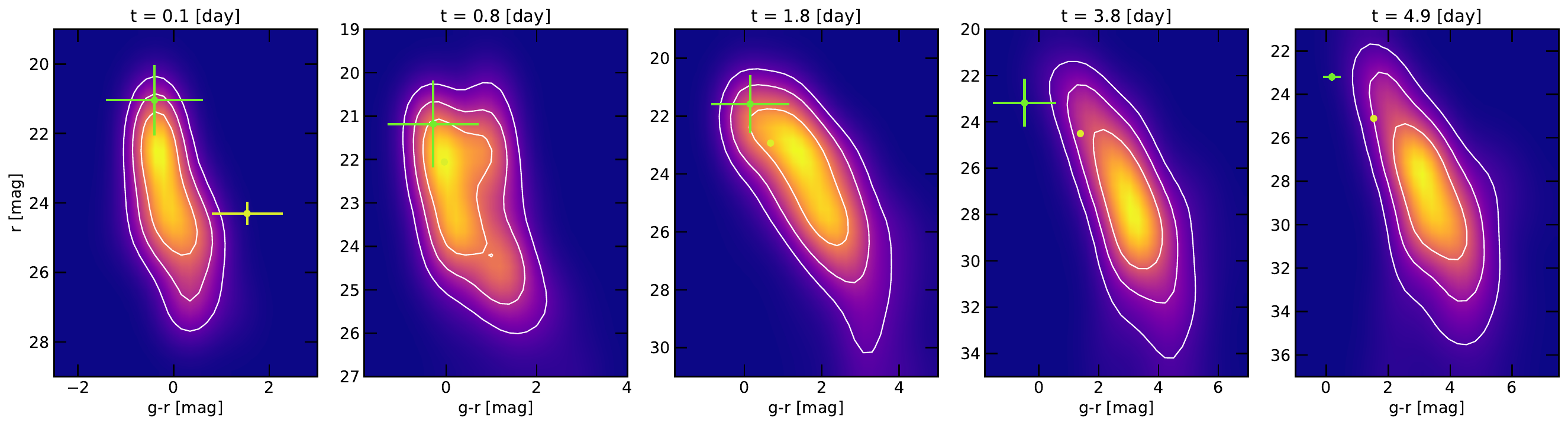}
    \caption{\textbf{The distribution of the r-band magnitudes and colors (g-r) of the optical transients (AT2025ulz in green and AT2017gfo in orange) against a comprehensive library of 97,200 theoretical kilonova models from \citet{wollaeger_2021_7335961} (see \citealt{2021ApJ...918...10W,2022PhRvR...4a3046R,2023PhRvR...5a3168K} for details). White counters indicate $1\sigma$, $2\sigma$ and $3\sigma$ confidence regions.
    AT2017gfo is shifted to $z = 0.08489$. The relative compatibility scores $S_{\rm KN}\left(\tfrac{\rm AT2025ulz}{\rm AT2017gfo }\right)$ of AT2025ulz and AT2017gfo with the kilonova model library are 0.81, 0.60, 0.18, 0.012 and 0.0022, corresponding to all observations available up to the epoch indicated in each panel, from left to right.}
    }
    \label{fig:knprior}
\end{figure*}

We factor in the information on the counterpart's age by assigning a score of 
$S_{\rm age}=1$ if archival upper limits point to a transient younger than 1 day, $S_{\rm age}=0.8$ for an event younger than 3 days, and $S_{\rm age}=0.5$ otherwise. 

The next property we consider is whether the observed photometry is consistent with a kilonova. 
We extract the r-band magnitudes and colors (g-r) from a total of 97,200 kilonova models \citep{wollaeger_2021_7335961} with grid points distributed in dynamic ejecta mass $m_d$ (0.001--0.1$M_\odot$), dynamic ejecta velocity $v_d$ (0.05--0.3$c$), wind ejecta mass $m_w$ (0.001--0.1$M_\odot$), wind ejecta velocity $v_w$ (0.05--0.3$c$), viewing angle (0--90 deg) and four morphology–composition families \cite[see][for details]{2021ApJ...918...10W,2022PhRvR...4a3046R,2023PhRvR...5a3168K}.
In Figure \ref{fig:knprior}, we compare the observations of the AT2025ulz and AT2017gfo against the comprehensive library of kilonova models. 
We compute the log-likelihood for each model based on all pre-epoch observational dataset D as $\ln\mathcal{L}_{i}(D)$, assuming the systematic error is $\sigma_{\rm m}=1$ mag related to the uncertainties in the kilonova models and an additional $\sigma_{\rm ground}=1$ mag uncertainty arising from host-subtraction limitations in ground-based imaging for AT2025ulz.
Under the assumption of a uniform prior over the grid, the Bayesian evidence for the entire model library (N$=$97,200) with the dataset D is
\begin{equation}
    Z(D) = \tfrac{1}{N}\textstyle\sum\nolimits_{i=1}^N \mathcal{L}_i(D).
\end{equation}
The relative compatibility of AT2025ulz and AT2017gfo with the kilonova model library is quantified through the ratio of their Bayesian evidences, i.e., Bayes factor,
\begin{equation}
    S_{\rm KN}\left(\tfrac{\rm AT2025ulz}{\rm AT2017gfo }\right)
= \min \left[1, \,\tfrac{Z({\rm AT2025ulz})}{Z({\rm AT2017gfo})}\right].
\end{equation}

We further quantify how the host–galaxy association and morphology affects 
the probability for an observed optical transient to be a kilonova. 
The probability of a chance coincidence, $P_{\rm cc}$, is evaluated in 
a standard fashion using \citet{Bloom2002} and is negligible for
both AT2025ulz and AT2017gfo.
Spiral galaxies host a wide variety of stellar transients,
however most of them can be distinguished from kilonovae based on their luminosity scale, color and temporal evolution at early times \citep{Barna2025}. This leaves CCSNe as the primary contamination class, since their early peaks, powered by shock–cooling emission, mimic kilonovae for several days.
Among the 1,858 CCSNe observed by ZTF between 2018 and 2024\footnote{\url{https://www.wis-tns.org/}}, 37 were flagged as kilonova impostors \citep{Barna2025}, but we identify only 3 cases 
consistent with KNe model grids before the SN bump. The resulting contamination rate in spiral galaxies is $R( {\rm KN}^c \mid {\rm Spiral}) = 10^5\times 3/1858\approx  161 \ {\rm Gpc^{-3}\ yr^{-1}}$ \citep{Perley2020}.
Considering spiral galaxies are more likely \citep[81\%; e.g.,][]{Chu2022} to host binary neutron star, we adopt the kilonova rate in spiral galaxies as \citep{Abbott2023} $R( {\rm KN} \mid {\rm Spiral}) \approx  194\ {\rm Gpc^{-3}\ yr^{-1}}$.
Using these rates, the probability that an observed transient in a spiral galaxy is a bona fide kilonova, conditioned only on its host morphology, is:
\begin{equation}
    P_{\rm galaxy}
= \frac{R({\rm KN}\!\mid\!{\rm Spiral})}
       {R({\rm KN}\!\mid\!{\rm Spiral})\!+\!R({\rm KN}^c\!\mid \!{\rm Spiral})}\approx 0.55
\end{equation}
Elliptical galaxies contain mainly old stellar populations. Consequently, under the simplifying assumption that $P_{\rm galaxy}=1$, we evaluate the relative kilonova likelihood for AT2025ulz, calibrated to AT2017gfo, based on its host morphology, was initially set to 1 when the host appeared elliptical, but drops to 0.55 when the spiral arm structures are identified from HST imaging.

\bibliographystyle{aasjournal}
\bibliography{bib}
\end{document}